\documentclass[sigconf,screen]{acmart}

\copyrightyear{2026}
\acmYear{2026}
\setcopyright{cc}
\setcctype{by}
\acmConference[HCOMP 2026]{2026 ACM Conference on Human-AI Complementarity and Alignment}{September 27--30, 2026}{Alexandria, VA, USA}
\acmBooktitle{2026 ACM Conference on Human-AI Complementarity and Alignment (HCOMP 2026), September 27--30, 2026, Alexandria, VA, USA}
\acmDOI{10.1145/3834580.3838746}
\acmISBN{979-8-4007-2894-5/2026/09}

\usepackage{graphicx}
\usepackage{booktabs}
\usepackage{array}    
\usepackage{pifont}
\usepackage{bm}

\usepackage{makecell}
\usepackage{caption}

\usepackage{xcolor}
\definecolor{customblue}{RGB}{0, 0, 255}

\usepackage{tabularx}
\usepackage{multirow} 
\usepackage{subcaption}
\usepackage{fancybox}
\usepackage{framed}

\usepackage{array}
\newcolumntype{L}[1]{>{\raggedright\arraybackslash}p{#1}}
\usepackage{placeins}

\usepackage{multirow}

\begin{document}

\title{Beyond Truth Discovery: A Two-Stage Framework to Assess \\ the Severity of False Claim during Disasters}


\author{Ruichen Yao}
\affiliation{%
  \institution{University of Illinois Urbana-Champaign}
  \city{Urbana}
  \country{USA}}
\email{ryao8@illinois.edu}

\author{Tejna Dasari}
\affiliation{%
  \institution{University of Illinois Urbana-Champaign}
  \city{Urbana}
  \country{USA}}
\email{tejnad2@illinois.edu}

\author{Gulshat Baispay}
\affiliation{%
  \institution{University of Illinois Urbana-Champaign}
  \city{Urbana}
  \country{USA}}
\affiliation{%
  \institution{Al-Farabi Kazakh National University}
  \city{Almaty}
  \country{Kazakhstan}}
\email{gbaispay@illinois.edu}

\author{Aizhan Zaurbek}
\affiliation{%
  \institution{University of Illinois Urbana-Champaign}
  \city{Urbana}
  \country{USA}}
\email{azaurbek@illinois.edu}

\author{Yifan Liu}
\affiliation{%
  \institution{University of Illinois Urbana-Champaign}
  \city{Urbana}
  \country{USA}}
\email{yifan40@illinois.edu}

\author{Yaokun Liu}
\affiliation{%
  \institution{University of Illinois Urbana-Champaign}
  \city{Urbana}
  \country{USA}}
\email{yaokunl2@illinois.edu}

\author{Zelin Li}
\affiliation{%
  \institution{University of Illinois Urbana-Champaign}
  \city{Urbana}
  \country{USA}}
\email{zelin3@illinois.edu}

\author{Dong Wang}
\affiliation{%
  \institution{University of Illinois Urbana-Champaign}
  \city{Urbana}
  \country{USA}}
\email{dwang24@illinois.edu}

\settopmatter{authorsperrow=4}

\renewcommand{\shortauthors}{Yao et al.}

\begin{abstract}
False information spreads rapidly on social media during disasters and can undermine emergency response efforts, public trust, and crisis communication. Existing research primarily focuses on determining whether social media posts contain false information, but provides limited insight into the specific false claims embedded within posts and the severity of individual false claims.
To address the limitations, we propose a two-stage framework to assess the severity of false claims during disasters. In the first stage, we develop a false claim extraction agent that identifies false claims from multimodal social media posts containing text, images, videos, and links. A subsequent verification step validates extracted claims with supporting evidence. In the second stage, we define false claim severity as the combination of two complementary dimensions: believability, which determines the likelihood that a claim will be believed, and harmfulness, which captures the potential consequences if it is believed.
Human annotators assess both dimensions to construct a claim-level severity benchmark using false claims extracted from Reddit posts related to hurricanes and wildfires.
Building upon this benchmark, we investigate false claim severity assessment as a human-AI alignment problem, evaluating whether models can reproduce human judgments under a shared evaluation rubric rather than merely predicting severity labels. Experiments on the benchmark show that traditional supervised models exhibit limited alignment with human judgments, whereas Large Language Models (LLMs) achieve substantially stronger performance. Among the evaluated strategies, in-context learning consistently achieves the strongest alignment with human judgments, highlighting the importance of human examples and shared decision criteria for severity assessment.
\end{abstract}

\ccsdesc[500]{Human-centered computing~Collaborative and social computing}
\ccsdesc[300]{Computing methodologies~Machine learning}
\ccsdesc[300]{Information systems~Information retrieval}
\ccsdesc[100]{Information systems~World Wide Web}



\keywords{False Claim Severity;
Human-AI Alignment;
Multi-modal Detection;
Believability;
Harmfulness;
Social Media;
Disaster
}



\maketitle
\section{Introduction}
\begin{figure*}[t]
    \centering
    \begin{minipage}{\textwidth} 
        \centering
        \includegraphics[width=0.9494\textwidth]{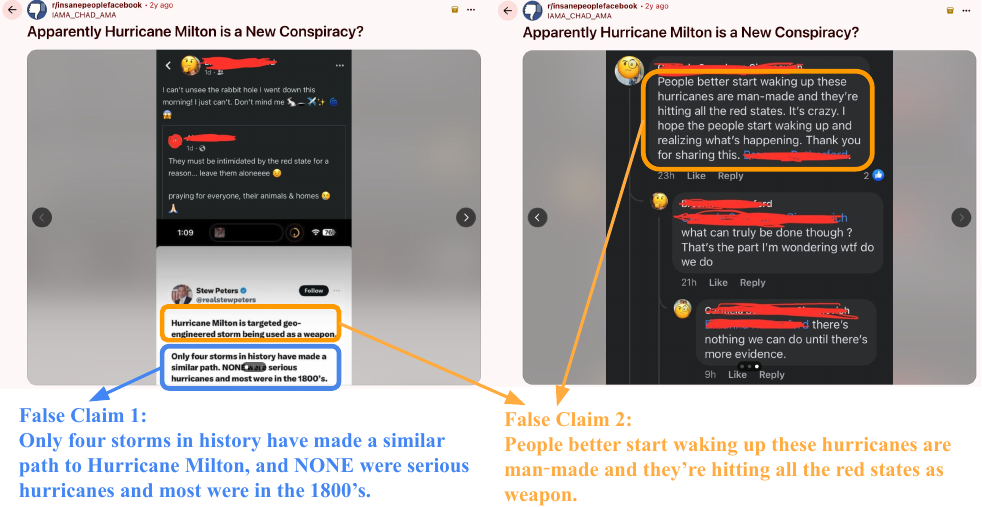} %
        \caption{Example of a Single Post Containing Multiple False Claims.}
        \label{fig:multi-claim-exp}
    \end{minipage}
\end{figure*}

The increasing frequency and severity of current natural disasters (e.g., hurricanes, wildfires, earthquakes, etc.) cause significant casualties and huge economic losses, leading to serious challenges to human society~\cite{mcentire2021disaster,yao2026mash,yao2026disimpact}. For example, in 2024, several strong hurricanes struck the United States, including Category 4 Hurricane Helene and Category 5 Hurricane Milton. The two hurricanes caused extensive damage, resulting in at least 250 casualties and over \$300 billion economic losses~\cite{ScienceHurricane, yao2026mash}. Social media serves as a critical communication channel during disasters, enabling millions of individuals to share and access information in real time~\cite{wang_social_sensing,marshall2016mood,yao2026mash,yao2026disimpact}.
By connecting affected communities, government agencies, and the broader public, social media facilitates the rapid exchange of disaster-related information across diverse stakeholders. The discussions on social media play an important role in shaping public perception, disseminating information, and supporting coordination among affected communities~\cite{Fighting_fake_news_during_disasters}.

However, false information can spread rapidly through social media and may lead to a variety of negative consequences, including inciting public panic, interfering with rescue operations, and undermining public trust in emergency response agencies ~\cite{Fighting_fake_news_during_disasters,Fighting_misinformationLatinAmerica}.
To prevent the spread of false claims, existing research designs truth discovery and veracity assessment algorithms to determine whether social media posts contain false information~\cite{wang2012scalability,liu2025modality}. The truth discovery models have achieved promising results in identifying false information on social media. However, existing approaches still suffer from two main limitations.

The first limitation is the lack of claim-level analysis of posts containing false information. A social media post may contain multiple pieces of information, while a claim corresponds to a specific assertion within the post that can be independently evaluated. Figure~\ref{fig:multi-claim-exp} presents an example in which a single social media post contains two distinct false claims. Although both claims originate from the same post, they convey different information and therefore require separate assessment. Existing false information detection approaches typically treat entire social media post as the holistic unit of analysis and classify the post as true or false~\cite{shang2025multitec,liu2025modality,gong2025designing}. 
By assigning a single label to the entire post, information about individual claims is compressed into a single representation. As a result, the post-level label cannot identify which specific claims are false and may overlook false claims embedded within the content.
This limitation becomes particularly evident when true and false information coexist within the same social media post~\cite{shu2017fake}. In these cases, a single post-level label cannot distinguish between true and false claims, nor can it capture the potential severity of individual false claims.
The limitation is further amplified in multimodal posts. False claims may be expressed not only through textual statements but also through images, videos, or combinations of multiple modalities. Although prior studies have explored mixed-truth false information~\cite{shu2017fake,thibault2025guide,guo2022survey} and multimodal false information detection~\cite{tahmasebi2024multimodal,liu2025modality,shen2024multimodal}, most approaches remain focused on determining the veracity of an entire post rather than explicitly identifying and analyzing individual false claims embedded within it.
This limitation highlights the importance of moving beyond post-level false information detection toward claim-level understanding, where individual false claims can be extracted and evaluated independently.

The second limitation is the lack of evaluation of the severity of false claims during crises.
Existing research primarily focuses on detecting false claims in social media posts but lacks an analysis of their severity~\cite{hurricane1,hurricane2, hurricane3, liu2025modality, shang2025multitec}.
Not all false claims lead to significant social consequences, as their severity varies based on content and audience perception.
For example, the claim ``Hurricanes are randomly given names without any organized process'' is false but unlikely to be believed and cause widespread panic or disrupt disaster relief efforts. On the other hand, false claims with high severity are more likely to be believed and cause panic and confusion among disaster survivors, significantly hindering rescue efforts. For instance, false claims like ``\$750 FEMA assistance offered to survivors is a loan and must be repaid'' could discourage victims from seeking financial aid, exacerbating their vulnerability during recovery. 
Understanding differences in false claim severity is critical because emergency agencies, fact-checking organizations, and social media platforms often face resource constraints that prevent them from addressing all false claims with equal urgency.
Assessing severity can support intervention prioritization by identifying false claims that are both highly believable and highly harmful, as these claims are most likely to influence public behavior and require immediate attention during emergencies.

To address the limitations of prior work, we propose a two-stage framework for false claim severity assessment during disasters. The first stage is to extract all false claims from the multimodal social media post. We adopt a Multimodal Large Language Model (MLLM) to analyze both textual and visual content within social media posts and compare the information against external online sources, including fact-checking websites, government reports, and reliable news articles, to identify and extract false claims. To improve reliability, we further introduce a self-verification step in which the extracted claims are validated against supporting evidence before being retained for downstream analysis.

Building upon the extracted claims, in the second stage we define false claim severity through two complementary dimensions: believability and harmfulness. Believability captures the extent to which affected audiences are likely to perceive a false claim as credible and trustworthy, whereas harmfulness reflects the potential consequences if the false claim were accepted as true. Using expert-designed annotation guidelines, human annotators assess both dimensions and establish a claim-level severity benchmark using false claims extracted from Reddit posts during hurricanes and wildfires. By jointly considering believability and harmfulness, the framework moves beyond traditional binary veracity assessment and organizes false claims into four severity categories.


Using this benchmark, we investigate false claim severity assessment as a human-AI alignment problem. Rather than treating severity assessment as a purely prediction task, we frame it as a human-AI alignment problem: can AI models reproduce the reasoning and decisions made by human annotators under a shared rubric? The severity of a false claim is inherently tied to human perceptions of believability and potential harm, therefore, meaningful assessment requires alignment with human judgment rather than merely generating predictions. 
We first evaluate traditional supervised models to examine whether conventional text classification approaches can reproduce human severity judgments. We then investigate LLM-based assessment using both proprietary and open models. To better understand how alignment can be improved, we further compare three assessment strategies: direct rubric-based assessment, reason-enhanced assessment, and in-context learning assessment by providing examples.
Experimental results show that traditional supervised models exhibit limited alignment with human severity judgments, whereas LLM-based approaches achieve substantially stronger performance. Among the assessment strategies, in-context learning consistently produces the best results, suggesting that exposure to human-labeled examples and shared decision criteria is more effective for achieving human-AI alignment than relying on annotation guidelines or explanations.
\section{Related Works}

\subsection{Disaster Truth Discovery on Social Media}
With the wide adoption of digital technology, social media has become one of the primary media for people to obtain and spread information during natural disasters.
False information is widely spread and have misled many people during hazards~\cite{Fighting_fake_news_during_disasters,Disaster_Misinformation_Management,hurricane1,hurricane2}. For example, \citet{Fighting_fake_news_during_disasters} indicate the prevalence of false information during natural disasters and terrorist attacks. They highlight the challenges posed by the rapid spread of false information on social media and discuss strategies to mitigate their impact~\cite{Fighting_fake_news_during_disasters}. \citet{Disaster_Misinformation_Management} explore various strategies emergency response agencies can employ to mitigate the impact of false information. \citet{hurricane1} and \citet{hurricane2} study the false information spread on X (previously known as Twitter) during the 2012 hurricane season. They point out the crucial role of verified accounts in curbing false information, as the public relies on their credible information to debunk false information~\cite{hurricane1,hurricane2}. In addition, a series of machine learning models have been proposed to detect false information in social media during natural disasters~\cite{Systematic_meta-analysis,2015detectfake,hurricane3}. 
However, from 2010 to 2023, 92\% of research on machine learning models for handling social media false information during crises and disasters is related to COVID-19~\cite{Systematic_meta-analysis}. Research on social media false information related to natural disasters (e.g., floods, earthquakes, hurricanes) is limited during this period of time. Between 2010 and 2019, only 9 peer-reviewed journal articles examined false information in the context of natural disasters~\cite{disaster_misinfo_review}, and from 2010 to 2023, research on disaster event false information
accounted for just 2\% of all studies on social media false information~\cite{Systematic_meta-analysis}. To address this gap, we conduct a large-scale study of false information spread during 2024 Pacific hurricanes and 2025 California wildfires.

\begin{figure*}[t]
    \centering
    \begin{minipage}{\textwidth} 
        \centering
        \includegraphics[width=0.9994\textwidth]{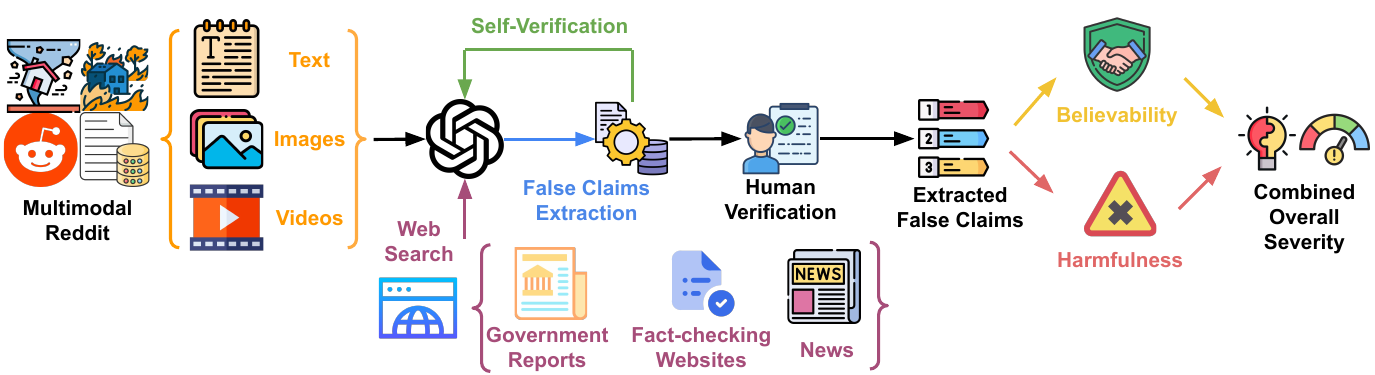} %
        \caption{Pipeline of the False Claim Severity Assessment.}
        \label{fig:pipeline}
    \end{minipage}
\end{figure*}

\subsection{In-Depth Research on False Information}

Research on social media veracity assessment emphasizes that existing research mainly focuses on identifying and verifying the existence of false information. Only a small amount of works focus on the risk, harm, and severity of false information~\cite{investigation,A_Tale_of_Two_crises,the_challenges,coviddrink}. For example, \citet{investigation} investigate the harms such as safety, panic, and exposure to diseases caused by false information spread through social media during humanitarian crises. In addition, \citet{A_Tale_of_Two_crises} analyze how false information during 2013 Boston Marathon Bombing and 2017 Oroville Dam Evacuation
exacerbate challenges in crisis management, hinder the delivery of aid, and affect public perceptions and behaviors. \citet{the_challenges} explore the implementation of new content moderation policies centered on the concept of harm to address the spread of false information during the pandemic. \citet{coviddrink} model the perceived risk of health-related false information and build a COVID-based dataset labeled according to the severity of its potential harm to human health.

Although prior studies highlight the importance of risk, harm, and societal consequences associated with false information, they do not provide a systematic framework for assessing the severity of individual false claims. Existing work primarily examines the consequences of false information at a conceptual or event level, whereas limited research has investigated the systematic severity assessment problem.
To address these gaps, we propose a two-stage framework for false claim severity assessment during disasters. We decompose false claim severity into two complementary dimensions, believability and harmfulness, and construct a human-annotated benchmark for claim-level severity assessment.


\section{False Claim Severity Assessment Framework}

In this paper, we propose a two-stage false claim severity assessment framework for false claims derived from social media. Figure~\ref{fig:pipeline} illustrates the overview pipeline of the framework. 
The first stage focuses on false claim extraction from the social media posts. We design an MLLM-based agent that analyzes multimodal social media posts and extracts individual false claims. We further invite three independent annotators to verify the extracted false claims to ensure the quality of the claims and the reliability of the agent.
The second stage focuses on false claim severity assessment. We invite annotators to independently evaluate sampled false claims along two complementary dimensions: believability and harmfulness. Believability measures the extent to which a false claim is likely to be trusted, while harmfulness evaluates the potential consequences if the claim is accepted. The combination of these two dimensions forms the overall severity of each false claim.

\subsection{False Claim Extraction}
The first stage of the framework focuses on extracting false claims from multimodal social media posts. To evaluate the proposed framework across different disaster contexts, we examine two recent large-scale disasters: the 2024 Pacific Hurricanes and the 2025 California Wildfires. For the hurricane case, we focus on two major hurricanes, \textit{Hurricane Helene} and \textit{Hurricane Milton}, that affected the United States in 2024. The combined damage caused by the two hurricanes is estimated to exceed \$300 billion~\cite{ScienceHurricane, yao2026mash}. For the wildfire case, we focus on the 2025 California Wildfires that impacted Los Angeles. According to NOAA Climate~\cite{Lindsey2025LAFires} and \citet{paglino2025excess}, the wildfires resulted in an estimated 440 deaths and the destruction of more than 15,000 homes.

We leverage two existing social media datasets, MASH~\cite{yao2026mash} and DisImpact~\cite{yao2026disimpact}, which comprise social media posts collected during the 2024 Pacific Hurricanes and the 2025 California Wildfires. Table~\ref{tab:CollectionNum} summarizes the subset of posts that were labeled as false information in the original datasets. Among the available social media platforms, we focus on Reddit because it provides rich multimodal content that combines text, images, videos, and external links within a single post, like the example in Figure~\ref{fig:multi-claim-exp}, making it particularly suitable for claim-level false claim extraction and severity assessment~\cite{TwitterandReddit,yao2026disimpact,yao2026mash}. As some posts had been removed or were no longer publicly accessible, we were able to obtain 521 available false information posts from the hurricane dataset and 390 from the wildfire dataset.


\subsubsection{False Claim Extraction Agent} \label{sec:false_claim_extraction}
To extract false claims from multimodal social media posts, we design a two-step MLLM-based agent framework consisting of an evidence-grounded extraction module and a self-verification module.
For each Reddit post, the agent first aggregates information from all available modalities, including textual content, images, videos, and external links. 
For video content, we follow the common practice in video–language research by extracting 10 representative frames as the keyframes~\cite{Howto100m,Videoclip,MiniGPT4-Video1}. Specifically, we uniformly sample candidate frames throughout the video and encode each frame using the CLIP vision encoder~\cite{radford2021learning} to obtain semantic image representations.
To reduce redundancy, we perform semantic keyframe selection based on CLIP embeddings, retaining a candidate frame only when its maximum cosine similarity to previously selected keyframes falls below a predefined threshold.
We also transcribe audio into text descriptions to provide more information about the video.
For external links, we retrieve and parse the webpage content to obtain the textual information. 
The text, images, video keyframes, audio transcripts, and external-link content are combined into a unified input context and provided to the MLLM for downstream false claim extraction.


Based on the aggregated context, the extraction agent first identifies assertions contained within the post and treats them as candidate claims. For each candidate claim, the agent generates search queries using the claim content and key entities mentioned in the post. These queries are submitted through the MLLM’s web-search functionality to retrieve evidence from reliable external sources, including government reports, fact-checking websites, and established news. The retrieved evidence is jointly analyzed with the original post content to determine whether a candidate claim is false, and claims identified as false are extracted together with the supporting evidence used to justify the assessment.

We employ GPT-5.1~\cite{openai_gpt51_blog} as the backbone MLLM. Since GPT-5.1 has a knowledge cutoff of September 2024, it possesses limited knowledge of disasters that we examined in this study. Therefore, we adopt GPT-5.1 to better simulate real-world disaster response scenarios, where neither human analysts nor AI systems possess complete prior knowledge of unfolding events and must rely on continuously updated external information.
Moreover, disaster-related information is dynamic and continuously evolving, with official casualty counts, evacuation orders, infrastructure damage reports, and emergency response policies frequently changing over time. Consequently, accurate false claim identification requires access to external evidence rather than relying solely on the model’s knowledge. The system prompt of the extraction step is available in Figure~\ref{fig:prompt-false-claim-extraction} in the Appendix.

The initial extraction step produces a set of candidate false claims together with the supporting evidence retrieved for each claim. However, these preliminary results may still contain errors, where factual claims are incorrectly identified as false claims due to retrieval errors or imperfect reasoning.
To improve extraction reliability, we introduce a self-verification step inspired by recent advances in agentic reflection and self-correction~\cite{shinn2023reflexion,kamoi2024can}. In this step, the MLLM receives both the extracted claim and its associated evidence and reassesses whether the evidence sufficiently supports the falsity judgment. Claims are retained only when the model determines that the retrieved evidence provides sufficient support for the falsity judgment. Claims that lack adequate evidential support are discarded.
This verification step serves as an additional quality-control mechanism, improving extraction reliability and reducing hallucination of the final false claim set.
The system prompt of the verification step is available in Figure~\ref{fig:prompt-False Claim Verification} in the Appendix. As shown in Table~\ref{tab:CollectionNum}, we extract 914 false claims from the hurricane case and 557 false claims from the wildfire case. On average, each false information post contains 1.75 false claims in the hurricane dataset and 1.43 false claims in the wildfire dataset, indicating that multiple false claims frequently coexist within a single post.

\begin{table}[t]
\centering
\small
\setlength{\tabcolsep}{2.5pt}
\caption{Distribution of False Posts and False Claims.}

\begin{tabular}{ccc}
\toprule
  & \textbf{Hurricanes} & \textbf{Wildfires}   \\
\midrule
Clean Reddit Posts in Datasets   & 12,301  & 9,456           \\
False Information Posts Identified in Datasets    & 694 & 471\\ 
Available False Information Posts & 521 & 390\\ 
\midrule
Extracted False Claims    & 914        & 557        \\ 
Average False Claims per Post & 1.7543        & 1.4282       \\ 
Sampled False Claims for Downstream Tasks  & 300        & 300       \\ 
\bottomrule
\end{tabular}

\label{tab:CollectionNum}
\end{table}

\subsubsection{False Claims Human Verification}
To evaluate the reliability of the false claim construction agent, we conducted a human verification. Specifically, three independent annotators manually reviewed a randomly sampled subset of 600 extracted false claims, consisting of 300 claims from the hurricane dataset and 300 claims from the wildfire dataset, to determine whether the extracted claims are false. All annotators completed institutional IRB training and received training on disaster-related false information and the annotation guidelines before participating in the annotation process.
To verify the agreement of annotations between human annotators, we adopted two metrics, Consistency and Fleiss’ Kappa $\kappa$ Score. Consistency calculates the proportion of samples that are completely consistent among all annotators, while Fleiss’ Kappa Score calculates the overall consistency after random consistency correction between annotators~\cite{fleiss1971measuring}.
Let $A$, $B$, and $C$ be the labels assigned by each annotator, 
$i$ represents the index of the data sample, and $N$ is the total number of data samples. 
The Consistency is defined as: 
$\text{Consistency} = \frac{\sum_{i=1}^{N} \mathbf(A_i = B_i = C_i)}{N}$.
Let $\bar{P}$ denote actual agreement (i.e., the average proportion of actual agreement between annotators) and $\bar{P_e}$ denote the expected agreement (i.e., the expected agreement if annotators randomly choose categories). The Fleiss' Kappa Score is defined as: 
$\text{Fleiss'} \ \kappa = \frac{\bar{P} - \bar{P}_e}{1 - \bar{P}_e}$.

Table~\ref{tab:agreement} shows the agreement between the three human annotators. The high inter-annotator agreement indicates that independent annotators consistently reached similar judgments when assessing whether an extracted claim constituted false or misleading information. We then aggregated the annotation results using majority voting and compared them with the predictions produced by the extraction agent. The extraction agent achieved an accuracy of 0.98 on the hurricane subset and 0.99 on the wildfire subset. The human verification results demonstrate the reliability of the proposed false claim extraction agent and suggest that the combination of evidence-grounded extraction and self-verification can effectively identify false claims from multimodal social media posts.

\begin{table}[t]
\centering
\small
\setlength{\tabcolsep}{2.5pt}
\caption{Annotation Agreement between Human Annotators.}

\begin{tabular}{ccc|cc}
\toprule
&
\multicolumn{2}{c|}{\textbf{Hurricanes}}
&
\multicolumn{2}{c}{\textbf{Wildfires}}
\\
\cmidrule(lr){2-3}
\cmidrule(lr){4-5}
&
Consistency & Fleiss' $\kappa$ & Consistency & Fleiss' $\kappa$ \\
\midrule

False Claim Verification & 0.9767 & 0.6738 & 0.9967 & 0.7989 \\


Believability Annotation & 0.7897 & 0.7100 & 0.8389 & 0.6575 \\

Harmfulness Annotation & 0.8241 & 0.7614 & 0.7550 & 0.6204 \\
\bottomrule
\end{tabular}

\label{tab:agreement}
\end{table}


\begin{figure*}[t]
    \centering
    \begin{minipage}{\textwidth} 
        \centering
        \includegraphics[width=0.94\textwidth]{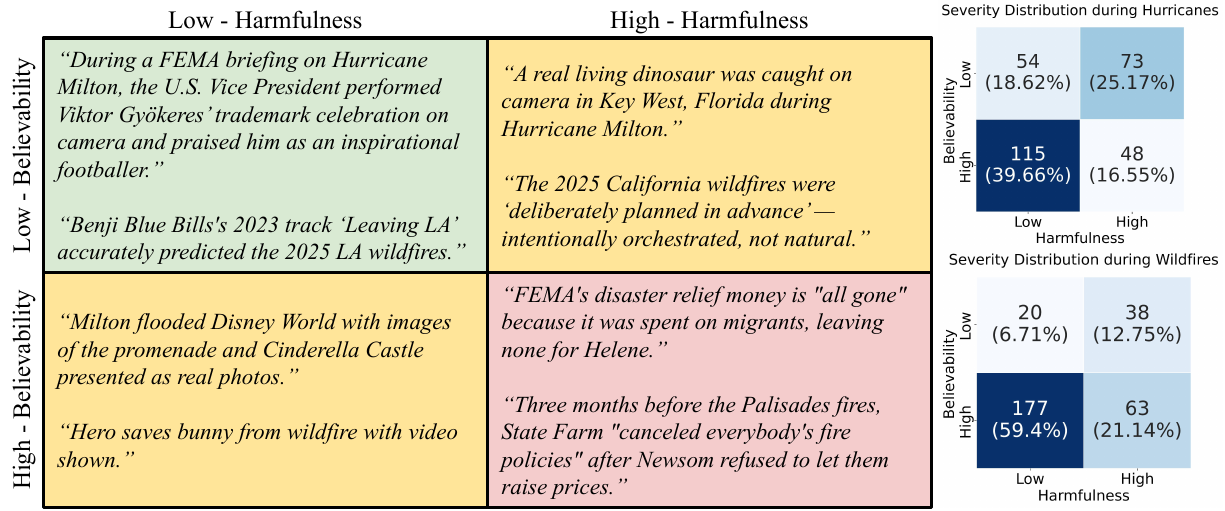} %
        \caption{Examples and distribution of false claims across the four severity categories derived from believability and harmfulness. Representative false claims are shown on the left, while the distribution of human gold standard is shown on the right.}
        \label{fig:severity-exp}
    \end{minipage}
\end{figure*}

\subsection{False Claim Severity Assessment} \label{sec:False Claim Severity Assessment}
In the second stage of the framework, we focus on assessing the severity of false claims. Different from factual verification, severity is inherently human-centered because it depends on how people perceive, interpret, and potentially act upon false information. Consequently, severity should not be treated as a single-dimensional construct because both the likelihood of belief and the consequences of claim acceptance jointly determine the risk posed by a false claim. Inspired by \citet{drolsbach2023believability} and \citet{sehat2024misinformation}, we conceptualize false claim severity as the combination of two complementary dimensions: believability, which captures the likelihood that a false claim will be perceived as credible and trustworthy, and harmfulness, which captures the potential risk that reliance on the claim could pose if it were believed. 

The importance of considering both dimensions can be illustrated through two contrasting examples. Some false claims may appear highly believable but lead to relatively limited consequences if believed. For example, a false claim that ``Hurricane Milton has flooded Disney World in Orlando, with images showing a flooded promenade around the castle caused by the storm'' may appear plausible within the context of hurricanes and therefore be perceived as believable. However, reliance on such a claim is unlikely to substantially affect high-risk consequences to the public.
In contrast, other false claims may imply severe consequences if believed as true but are too implausible to be believed by most audiences. For example, conspiracy claims like  ``The destruction of Disney World by Hurricane Milton has allowed the revelation of extensive underground tunnels used for human trafficking and other horrible crimes'' could potentially undermine trust in public institutions and contribute to social unrest if believed. However, the extraordinary nature of such claims may substantially reduce their perceived credibility. Therefore, a comprehensive assessment of severity requires jointly considering both dimensions, allowing false claims to be differentiated according to their believability and harmfulness.

\subsubsection{Believability Annotation}
Believability measures the extent to which a false claim is likely to be perceived as credible and trustworthy by affected audiences. We treat believability as a binary construct consisting of low and high believability, where highly believable claims are more likely to be accepted as true. This binary formulation provides a practical level of granularity while reducing annotation ambiguity and improving annotation consistency. It also helps mitigate data sparsity when constructing severity categories from combinations of believability and harmfulness.
Specifically, the same annotators evaluate whether the false claim describes a situation that could realistically occur within a disaster context. A false claim is classified as having high believability if it resembles realistic disaster events or updates, aligns with ordinary disaster experiences, or appears commonly encountered in disaster communications. In contrast, a false claim is classified as low believability if it depends on extraordinary conspiracy theories, describes unrealistic secret operations, or requires implausible capabilities.

To construct a human-centered severity benchmark, we invited the same three annotators to assess the believability of extracted false claims. Following the same sampling strategy used in the false claim verification stage, we randomly selected 300 false claims from the hurricane dataset and 300 false claims from the wildfire dataset, resulting in a total of 600 claims for annotation. Each annotator independently reviewed the claims according to a shared annotation rubric. Table~\ref{tab:agreement} shows high inter-annotator agreement results for the believability annotation task. The detailed rubric of believability annotation is presented in Table~\ref{tab:believability_rubric} in the Appendix.

\subsubsection{Harmfulness Annotation}
Harmfulness measures the extent to which a false claim could lead to adverse consequences if it is accepted as true. Within our severity framework, harmfulness captures the potential risk associated with reliance on a false claim after belief has already occurred. Similar to believability, we also treat harmfulness as a binary construct consisting of low-risk consequences and high-risk consequences, where highly harmful claims are more likely to negatively affect disaster response, public safety, or community well-being. This formulation enables consistent annotation of potential disaster-related harms while avoiding subjective distinctions regarding the exact magnitude of consequences. It also supports the construction of interpretable severity categories when combined with believability.
Specifically, the same annotators are instructed to assume that affected audiences already believe the false claim and then evaluate the potential consequences that could result from such belief. A false claim is classified as having high-risk consequences if believing the claim could plausibly alter disaster-related behavior or coordination that negatively affects public safety, evacuation, medical response, or access to essential resources. 
In contrast, a false claim is classified as having low-risk consequences if belief in the claim is unlikely to produce substantial effects on disaster-related decisions, public safety, emergency operations, or community recovery.

We invited the same three annotators to independently assess the harmfulness of extracted false claims. Following the same sampling strategy used in the believability annotation task, we adopted the same randomly selected 300 false claims from the hurricane dataset and 300 false claims from the wildfire dataset. Each annotator independently reviewed the claims according to a shared annotation rubric. Table~\ref{tab:agreement} shows high inter-annotator agreement for the harmfulness annotation task. The detailed harmfulness annotation rubric is provided in Table~\ref{tab:harmfulness_rubric} in the Appendix.

\begin{table*}[t]
\centering
\small

\caption{Experiment Results on Traditional Supervised Models.}
\label{tab:finetune_exp}

\begin{tabular}{ccccc|ccc|ccc}
\toprule[1.5pt]

&
&
\multicolumn{3}{c|}{\textbf{Believability}}
&
\multicolumn{3}{c|}{\textbf{Harmfulness}}
&
\multicolumn{3}{c}{\textbf{Overall Severity}}

\\

\cmidrule(lr){3-5}
\cmidrule(lr){6-8}
\cmidrule(lr){9-11}

&
&
Accuracy & F1 & Cohen's $\kappa$
&
Accuracy & F1 & Cohen's $\kappa$
&
Accuracy & F1 & Cohen's $\kappa$
\\

\midrule

\multirow{4}{*}{\rotatebox[origin=c]{90}{\textbf{Hurricanes}}}
& RoBERTa
& $0.51 \pm 0.07$ & $0.44 \pm 0.09$ & $0.02 \pm 0.08$
& $0.53 \pm 0.05$ & $0.41 \pm 0.05$ &  $0.03 \pm 0.03$
& $0.28 \pm 0.07$ & $0.19 \pm 0.06$ & $0.02 \pm 0.04$
\\

& BART
& $0.53 \pm 0.04$ & $0.51 \pm 0.04$ & $0.04 \pm 0.09$
& $0.51 \pm 0.04$ & $0.49 \pm 0.03$ & $0.03 \pm 0.07$
& $0.29 \pm 0.02$ & $0.24 \pm 0.01$ & $0.03 \pm 0.03$
\\

& ELECTRA
& $0.54 \pm 0.03$ & $0.52 \pm 0.03$ & $0.05 \pm 0.06$
& $0.48 \pm 0.04$ & $0.48 \pm 0.04$ & $0.01 \pm 0.07$
& $0.27 \pm 0.03$ & $0.24 \pm 0.03$ & $0.02 \pm 0.05$
\\

& ConvBERT 
& $0.54 \pm 0.04$ & $0.53 \pm 0.03$ & $0.06 \pm 0.06$
& $0.57 \pm 0.06$ & $0.56 \pm 0.06$ & $0.14 \pm 0.13$
& $\textbf{0.34} \pm \textbf{0.06}$ & $\textbf{0.30} \pm \textbf{0.05}$ & $\textbf{0.09} \pm \textbf{0.08}$
\\

\midrule

\multirow{4}{*}{\rotatebox[origin=c]{90}{\textbf{Wildfires}}}
& RoBERTa
& $0.75 \pm 0.05$ & $0.55 \pm 0.08$ & $0.15 \pm 0.13$
& $0.54 \pm 0.05$ & $0.47 \pm 0.05$ & $0.02 \pm 0.08$
& $0.40 \pm 0.06$ & $0.24 \pm 0.03$ & $0.05 \pm 0.04$
\\

& BART
& $0.67 \pm 0.09$ & $0.55 \pm 0.04$ & $0.14 \pm 0.06$
& $0.54 \pm 0.07$ & $0.49 \pm 0.03$ & $0.01 \pm 0.04$
& $0.37 \pm 0.07$ & $0.28 \pm 0.05$ & $0.04 \pm 0.03$
\\

& ELECTRA
& $0.79 \pm 0.04$ & $0.59 \pm 0.08$ & $0.20 \pm 0.15$
& $0.55 \pm 0.06$ & $0.51 \pm 0.05$ & $0.02 \pm 0.09$
& $0.44 \pm 0.04$ & $0.27 \pm 0.05$ & $0.05 \pm 0.05$
\\

& ConvBERT 
& $0.74 \pm 0.05$ & $0.54 \pm 0.08$ & $0.10 \pm 0.16$
& $0.60 \pm 0.03$ & $0.57 \pm 0.04$ & $0.14 \pm 0.07$
& $\textbf{0.50} \pm \textbf{0.03}$ & $\textbf{0.36} \pm \textbf{0.07}$ & $\textbf{0.14} \pm \textbf{0.04}$
\\

\midrule

\multirow{4}{*}{\rotatebox[origin=c]{90}{\textbf{Cross}}}
& RoBERTa
& $0.64 \pm 0.02$ & $0.61 \pm 0.03$ & $0.23 \pm 0.06$
& $0.56 \pm 0.06$ & $0.46 \pm 0.05$ & $0.01 \pm 0.09$
& $0.38 \pm 0.05$ & $0.29 \pm 0.04$ & $0.10 \pm 0.04$
\\

& BART
& $0.63 \pm 0.03$ & $0.59 \pm 0.04$ & $0.19 \pm 0.08$
& $0.53 \pm 0.05$ & $0.50 \pm 0.03$ & $0.07 \pm 0.03$
& $0.36 \pm 0.02$ & $\textbf{0.30} \pm \textbf{0.03}$ & $\textbf{0.11} \pm \textbf{0.04}$
\\

& ELECTRA
& $0.65 \pm 0.03$ & $0.55 \pm 0.05$ & $0.11 \pm 0.09$
& $0.54 \pm 0.04$ & $0.52 \pm 0.04$ & $0.05 \pm 0.07$
& $0.37 \pm 0.03$ & $0.28 \pm 0.03$ & $0.06 \pm 0.04$
\\

& ConvBERT 
& $0.64 \pm 0.01$ & $0.53 \pm 0.02$ & $0.07 \pm 0.04$
& $0.57 \pm 0.03$ & $0.54 \pm 0.02$ & $0.09 \pm 0.05$
& $\textbf{0.39} \pm \textbf{0.02}$ & $0.28 \pm 0.02$ & $0.08 \pm 0.03$
\\

\bottomrule[1.5pt]

\end{tabular}

\end{table*}

\subsubsection{Severity Construction}
For both the believability and harmfulness dimensions, we aggregate the annotations from the three annotators using majority voting to construct the human reference labels. The resulting labels serve as the human gold standard for subsequent severity assessment and human-AI alignment analysis.

We then combine the believability and harmfulness labels to construct the final severity representation of each false claim. Since severity is jointly determined by the believability and harmfulness, the combination of the two binary dimensions yields four possible severity categories. Specifically, a false claim may be characterized as: (1) low believability and low harmfulness, (2) high believability and low harmfulness, (3) low believability and high harmfulness, or (4) high believability and high harmfulness. Figure~\ref{fig:severity-exp} illustrates the examples of each severity category and the distribution of human gold standard after the annotation.
Among these categories, false claims exhibiting both high believability and high harmfulness represent the highest level of severity, as they are simultaneously more likely to be accepted as true and more likely to produce adverse consequences. In contrast, false claims with low believability and low harmfulness represent the lowest level of severity.
\begin{table*}[t]
\centering
\small

\caption{Experiment Results on LLM-based Models.}
\label{tab:exp}

\begin{tabular}{ccccc|ccc|ccc}
\toprule[1.5pt]

&
&
\multicolumn{3}{c|}{\textbf{Believability}}
&
\multicolumn{3}{c|}{\textbf{Harmfulness}}
&
\multicolumn{3}{c}{\textbf{Overall Severity}}

\\

\cmidrule(lr){3-5}
\cmidrule(lr){6-8}
\cmidrule(lr){9-11}

&
&
Accuracy & F1 & Cohen's $\kappa$
&
Accuracy & F1 & Cohen's $\kappa$
&
Accuracy & F1 & Cohen's $\kappa$
\\

\midrule

\multirow{6}{*}{\rotatebox[origin=c]{90}{\textbf{Hurricanes}}}
& GPT-5.1 Direct
& 0.9000 & 0.8997 & 0.8002
& 0.6793 & 0.6728 & 0.4029
& 0.6069 & 0.5857 & 0.4856
\\

& GPT-5.1 Reasoned
& 0.8966 & 0.8956 & 0.7913
& 0.6655 & 0.6557 & 0.3825
& 0.5966 & 0.5634 & 0.4695
\\

& GPT-5.1 In-Context
& 0.9069 & 0.9066 & 0.8140
& 0.7379 & 0.7358 & 0.5042
& \textbf{0.6759} & \textbf{0.6549} & \textbf{0.5679}
\\
\cmidrule(lr){2-11}

& Qwen3-32B Direct
& 0.7379 & 0.7338 & 0.4990
& 0.7069 & 0.7068 & 0.4262
& 0.5069 & 0.4892 & 0.3571
\\

& Qwen3-32B Reasoned
& 0.7655 & 0.7633 & 0.5488
& 0.6966 & 0.6965 & 0.4119
& 0.5276 & 0.5109 & 0.3827
\\

& Qwen3-32B In-Context
& 0.7621 & 0.7593 & 0.5433
& 0.7448 & 0.7448 & 0.5022
& \textbf{0.5655} & \textbf{0.5469} & \textbf{0.4265}
\\

\midrule[1.2pt]

\multirow{6}{*}{\rotatebox[origin=c]{90}{\textbf{Wildfires}}}
& GPT-5.1 Direct
& 0.8926 & 0.8351 & 0.6704
& 0.7181 & 0.7129 & 0.4458
& 0.6409 & 0.6051 & 0.4546
\\

& GPT-5.1 Reasoned
& 0.8993 & 0.8137 & 0.6306
& 0.7148 & 0.7112 & 0.4474
& 0.6342 & 0.6020 & 0.4382
\\

& GPT-5.1 In-Context
& 0.9161 & 0.8671 & 0.7341
& 0.7114 & 0.7095 & 0.4513
& \textbf{0.6678} & \textbf{0.6365} & \textbf{0.4930}
\\
\cmidrule(lr){2-11}
& Qwen3-32B Direct
& 0.6812 & 0.6468 & 0.3571
& 0.6946 & 0.6839 & 0.3799
& 0.4463 & 0.3905 & 0.2454
\\

& Qwen3-32B Reasoned
& 0.7248 & 0.6796 & 0.3978
& 0.6913 & 0.6844 & 0.3878
& 0.4799 & 0.4261 & 0.2778
\\

& Qwen3-32B In-Context
& 0.6846 & 0.6513 & 0.3665
& 0.7651 & 0.7497 & 0.5022
& \textbf{0.5235} & \textbf{0.4430} & \textbf{0.3200}
\\

\midrule[1.2pt]

\multirow{6}{*}{\rotatebox[origin=c]{90}{\textbf{Cross}}}
& GPT-5.1 Direct
& 0.8963 & 0.8837 & 0.7679
& 0.6990 & 0.6990 & 0.4325
& 0.6241 & 0.6036 & 0.4866
\\

& GPT-5.1 Reasoned
& 0.8980 & 0.8795 & 0.7592
& 0.6905 & 0.6902 & 0.4232
& 0.6156 & 0.5909 & 0.4720
\\

& GPT-5.1 In-Context
& 0.9116 & 0.9003 & 0.8008
& 0.7245 & 0.7245 & 0.4777
& \textbf{0.6718} & \textbf{0.6502} & \textbf{0.5455}
\\
\cmidrule(lr){2-11}
& Qwen3-32B Direct
& 0.7092 & 0.7064 & 0.4542
& 0.7007 & 0.6975 & 0.4073
& 0.4762 & 0.4605 & 0.3148
\\

& Qwen3-32B Reasoned
& 0.7449 & 0.7396 & 0.5054
& 0.6939 & 0.6925 & 0.4037
& 0.5034 & 0.4915 & 0.3453
\\

& Qwen3-32B In-Context
& 0.7228 & 0.7198 & 0.4772
& 0.7551 & 0.7509 & 0.5086
& \textbf{0.5442} & \textbf{0.5140} & \textbf{0.3875}
\\

\bottomrule[1.5pt]

\end{tabular}

\end{table*}

\section{Human-AI Alignment}

\subsection{Experimental Setup}

To investigate how different modeling approaches can align with human judgments on false claim severity detection, we conduct experiments on the 600 false claims that were manually annotated by three human annotators. The majority-voted human annotations described in Section~\ref{sec:False Claim Severity Assessment} serve as the gold standard for all alignment evaluations. Following the proposed severity framework, all models independently assess the believability and harmfulness of each false claim using the same annotation rubric provided to human annotators. 
The resulting predictions are then combined to determine the final severity category for each false claim. This formulation yields four severity categories: (1) low believability and low harmfulness, (2) high believability and low harmfulness, (3) low believability and high harmfulness, or (4) high believability and high harmfulness.

To examine the robustness of human-AI alignment across disaster contexts, we conduct experiments under three dataset settings: Hurricane-only, Wildfire-only, and Cross-disaster (Hurricane and Wildfire). The Hurricane-only and Wildfire-only settings evaluate alignment within a single disaster domain, whereas the Cross-disaster setting combines false claims from both disasters into a unified dataset. These settings allow us to investigate whether alignment patterns remain consistent across different disaster contexts and under a more diverse multi-disaster environment. The Cross-disaster setting is designed to evaluate the robustness and generalizability of severity assessment across different disasters. This setting allows us to examine whether the proposed severity framework and human-AI alignment patterns remain consistent when models are exposed to broader false claims. As shown in Figure~\ref{fig:severity-exp}, the severity categories are unevenly distributed in both the hurricane and wildfire datasets. To comprehensively evaluate model performance and human-AI alignment under class imbalance, we report Accuracy, Macro-F1, and Cohen’s Kappa.
Accuracy measures the overall performance of the model. We additionally report Macro-F1, which gives equal weight to all classes and better reflects performance on minority severity categories~\cite{yao2026disimpact,yao2026mash}. In addition, we report Cohen’s Kappa as it quantifies the agreement between two raters by correcting for the proportion of agreements that would be expected purely by chance, making it more appropriate for assessment in human-AI alignment task~\cite{cohen1960coefficient}.

\subsection{Traditional Supervised Models}

To establish a baseline for human-AI alignment in false claim severity assessment, we first investigate whether traditional supervised NLP models can reproduce human judgments on false claim severity detection. Specifically, we select RoBERTa~\cite{roberta}, BART~\cite{bart}, ELECTRA~\cite{electra}, and ConvBERT~\cite{ConvBERT} because of their strong performance on a wide range of tasks, as well as their popularity and adoption as standard baselines in prior works~\cite{mashkoor2025comparative, yao2026mash}.
For each model, we perform 5-fold cross-validation to obtain a robust performance. In each fold, models are trained for 10 epochs and evaluated on the corresponding test fold. To account for variability across different train-test splits, we report the mean and standard deviation of all evaluation metrics across the five folds.

The experiments are conducted on NVIDIA L40 GPU, and Table~\ref{tab:finetune_exp} presents the performance of traditional supervised models across different dataset settings. For the individual believability and harmfulness tasks, most models achieve moderate accuracy. However, performance declines when the two dimensions are combined into the final severity categories, indicating that accurately reproducing human severity judgments is substantially more challenging than predicting either dimension in isolation. Severity assessment involves four categories derived from combinations of believability and harmfulness, resulting in a more complex classification problem. In addition, F1 and Cohen’s Kappa scores are consistently low across all settings. This discrepancy suggests that model performance is largely driven by majority classes, while performance on minority severity categories remains limited. The low Cohen’s Kappa scores further indicate that model predictions do not consistently align with human judgments after accounting for agreement that could occur by chance.

The gap between Accuracy and the corresponding F1 and Cohen’s Kappa scores suggests that model performance is not uniformly distributed across severity categories. While traditional supervised models can correctly classify a proportion of instances, their performance decreases when evaluated across minority categories and when alignment with human annotations is explicitly considered.
These results highlight the difficulty of false claim severity assessment beyond conventional text classification tasks, suggesting that relying on semantic representations of the false claim is insufficient for achieving strong alignment with humans.

\subsection{LLM-based Models}

The limited performance of traditional supervised models suggests that false claim severity assessment cannot be effectively addressed through conventional text classification approaches alone. Therefore, we investigate whether LLMs can better reproduce human judgments of false claim severity.
We evaluate two representative LLMs, GPT-5.1~\cite{openai_gpt51_blog}, a frontier model with reasoning effort, and Qwen3-32B~\cite{qwen3technicalreport}, a widely used open-weight model. This selection allows us to compare human-AI alignment across both proprietary and open-weight LLMs with different model scales. To ensure reproducibility and minimize output variability, all experiments are conducted with the temperature set to 0. Following the human annotation framework, both models assess false claims under the same annotation rubric provided to human annotators. Specifically, we adopt three assessment strategies, \textit{Direct Assessment}, \textit{Reasoned Assessment}, and \textit{In-Context Assessment}. Direct Assessment requires the model to directly output the corresponding label. Reasoned Assessment requires the model to provide both a label and an accompanying explanation. In-Context Assessment augments the prompt with human-annotated examples that illustrate both positive and negative cases before the model performs the assessment. The system prompts are present in Figures~\ref{fig:prompt-Believability} and~\ref{fig:prompt-Harmfulness} in the Appendix.

The experiments are conducted on NVIDIA L40 GPU, and Table~\ref{tab:exp} presents the results of LLM-based severity assessment. In both hurricane dataset and wildfire dataset, LLM-based approaches substantially outperform the traditional supervised models. The improvements are particularly evident in F1 and Cohen’s Kappa, indicating stronger alignment with human annotations across believability, harmfulness, and overall severity.
The impact of the assessment strategy differs across models. For GPT, Reasoned Assessment produces comparable performance to Direct Assessment, suggesting that explicitly generating explanations provides limited benefit. In contrast, Qwen consistently improves under Reasoned Assessment, indicating that explanation generation may be particularly helpful for models with a smaller parameter size. One possible explanation is that larger frontier models already internalize much of the reasoning required for severity assessment, whereas smaller models benefit from explicitly organizing their reasoning process.

Among all strategies, In-Context Assessment consistently achieves the strongest performance for both GPT and Qwen, with particularly noticeable improvements on the overall severity assessment task. These results suggest that providing human-annotated examples is more effective than simply encouraging additional reasoning. In other words, exposure to representative human judgments appears to provide a stronger alignment signal than generating free-form explanations alone.
Despite these improvements, a noticeable gap remains between LLM predictions and the majority-voted human annotations. This result suggests that reproducing human assessments of believability and harmfulness remains a challenging task. Further research is needed to develop more effective approaches for aligning model predictions with human judgments.
\section{Limitations}
Our study has several limitations. Firstly, although human verification confirms that the extracted false claims are highly accurate, this validation only assesses the correctness of the extracted claims rather than the recall of the extraction. Manually identifying all possible false claims within multimodal social media posts is prohibitively expensive, we cannot determine whether additional valid claims were missed. Future work could investigate more comprehensive recall-oriented evaluation protocols using exhaustively annotated subsets or multiple independent extraction systems.

Secondly, the relatively small size of the annotated data may have affected the performance of the supervised models, as limited training data makes it more difficult to learn the patterns of false claim severity. Therefore, the observed performance gap should be interpreted with this limitation in mind. Under the current data setting, LLM-based approaches demonstrate stronger alignment with human judgments. Future work with larger and more diverse severity-annotated data will enable a more comprehensive comparison between supervised models and LLM-based approaches.

\section{Conclusion}

In this paper, we presented a two-stage false claim severity assessment framework in disaster-related social media. Different from prior works that primarily focus on post-level false information detection, we move toward claim-level analysis by introducing a multimodal false claim extraction agent capable of identifying individual false claims from textual and visual content. Human verification results demonstrate that the extraction agent can reliably construct a high-quality false claim benchmark from complex multimodal social media posts.
Building upon the extracted claims, we introduced a novel severity assessment framework that characterizes false claim severity through two complementary dimensions: believability and harmfulness. We conducted large-scale human annotation of false claims collected from hurricanes and wildfires and observed strong agreement among annotators. By explicitly separating the likelihood that a false claim will be believed from the consequences that may arise if it is believed, our framework enables a comprehensive understanding of the false claim.
Finally, we investigated how human-AI alignment can be leveraged to enable models to reproduce human judgments of false claim severity. Experimental results show that traditional supervised models exhibit limited alignment with human judgments, whereas LLM-based approaches achieve substantially stronger performance. Among the evaluated strategies, in-context learning consistently produces the best results, suggesting that exposure to human-labeled examples and shared decision criteria provides a stronger alignment signal than additional reasoning alone.
We hope that this work provides a foundation for future research on false claim severity assessment, human-AI alignment, and scalable moderation of disaster-related false information.
\bibliographystyle{ACM-Reference-Format}
\bibliography{sample-base}

\appendix

\begin{figure*}[!ht]
\centering
\footnotesize
\setlength{\fboxsep}{8pt}
\ovalbox{%
\begin{minipage}{0.95\textwidth} 
You are reviewing a multimodal social media post. The post may contain:

- publish time of the post

- title and text description

- images attached to the post (treat them as visual content reference, and extract any words in the images as part of the content)

- keyframes of video (treat them as visual content reference, and extract any words in the images as part of the content)

- video transcription text (treat it as spoken content from the video)

- text extracted from attached external links

Carefully analyze all input modalities together, including both the text, images, video, and transcription. If there are words in the images, you need to extract all words and read them as part of the content.

Your task is NOT to classify the entire post as true or false. Instead, your task is to identify ALL clear false or misleading claims contained in the post. False or misleading claims should still be extracted if the post quotes, references, discusses, repeats false claims made by another person or source, the claim itself should still be extracted and verified.

Important:

- A post may contain both true and false information.

- A post may contain multiple false claims.

- Only extract all claims that are false or misleading.

For each extracted false claim:

1. Provide the complete description of the false claim in the post. Comprehensively desceribe the claim in detail, including all relevant information such as numbers, locations, dates, and any specific assertions made in the post. Don't hulluanize the content that doesn't describe in the post.

2. Explain why the claim is false or misleading using reliable external evidence. 

When verifying claims, prioritize reliable sources such as:

- official government agencies

- official press releases or public documents

- verified fact-checking websites

- major reliable news organizations

For each claim, include:

- a concise explanation of why the claim is false

- supporting evidence from reliable sources

- source URLs whenever available

If no clear false claim exists in the post, return an empty list.

Respond in strict JSON format:

[

    \{
      "claim": "The complete description of the false claim. Do not hallucinate any content that is not described in the post. Don't explain why the claim is false in this field, just describe the claim itself.",
      
      "explanation": "Why the claim is false or misleading, supported by reliable evidence."
    \},
    
    ... list more if there are more false claims
    
]
\end{minipage}%

}

\caption{System Prompt for False Claim Extraction.}

\label{fig:prompt-false-claim-extraction}

\end{figure*}


\begin{figure*}[h!]
\centering
\footnotesize
\setlength{\fboxsep}{8pt}
\ovalbox{%
\begin{minipage}{0.95\textwidth} 
You are verifying whether an extracted claim from a media post is actually a false or misleading claim.

The input contains two parts:

1. An extracted false claim from a social media post.

2. Evidence or reference information related to the extracted falseclaim.

Your task is to verify whether the extracted false claim should truly be considered a false or misleading claim based on the provided evidence.

Carefully evaluate:

- whether the claim contradicts reliable evidence,

- whether the claim contains fabricated, inaccurate, misleading, or unsupported information,

- whether the evidence sufficiently supports labeling the claim as false or misleading.

A claim should be labeled as false (judgement=true) if:

- reliable evidence clearly shows the claim is factually incorrect,

- misleading in a verifiable way,

- fabricated,

- or substantially unsupported by available evidence.

Return your answer in strict JSON format matching the schema:

\{
  "judgement": true or false,
  
  "explanation": "Concise explanation of the judgement supported by the provided evidence in English."
\}
\end{minipage}%

}
\caption{System Prompt for False Claim Verification.}
\label{fig:prompt-False Claim Verification}

\end{figure*}


\begin{table*}[h!]
\centering
\caption{Believability Annotation Rubric.}
\label{tab:believability_rubric}
\small
\begin{tabular}{p{1cm} p{6cm} p{8cm}}
\toprule
\textbf{Label} & \textbf{Criteria} & \textbf{Example} \\
\midrule

\textbf{2}
&
\begin{itemize}
\item Resembles realistic disaster events or updates
\item Aligns with ordinary disaster experiences
\item Uses plausible real-world explanations
\item Appears similar to information commonly seen in disaster communication, reporting, or public discussion
\end{itemize}
&
\textit{``Hurricane Milton has flooded Disney World in Orlando, with images circulating online that supposedly show a flooded promenade at Walt Disney World's Cinderella Castle caused by the storm.''}

The claim resembles realistic hurricane-related flooding scenarios, references real locations and recognizable landmarks, and appears similar to information commonly shared during disasters.
\\

\midrule

\textbf{1}
&
\begin{itemize}
\item Depends on extraordinary conspiracies
\item Describes unrealistic secret operations
\item Makes fantastical accusations
\item Requires implausible scientific or technological capabilities
\end{itemize}
&
\textit{``The destruction of Disney World by Hurricane Milton has allowed or will allow the revelation of extensive underground tunnels beneath Disney used for human trafficking of children and other horrible crimes, supposedly being cleared out by the U.S. military.''}

The claim relies on extraordinary conspiratorial assumptions, including secret trafficking operations and covert military cleanup activities, which fall far outside realistic disaster communication.
\\

\bottomrule
\end{tabular}
\end{table*}


\begin{table*}[h!]
\centering
\caption{Harmfulness Annotation Rubric.}
\label{tab:harmfulness_rubric}
\small
\begin{tabular}{p{1cm} p{6cm} p{8cm}}
\toprule
\textbf{Label} & \textbf{Criteria} & \textbf{Example} \\
\midrule

\textbf{2}
&
Assume people already believe the false claim, then determine whether the consequence is a high-risk consequence or low-risk consequence.

A consequence should be considered highly harmful if believing the false claim could plausibly alter disaster-related behavior, decision-making, or coordination in:
\begin{itemize}
\item public safety (casualties, deaths)
\item evacuation or sheltering
\item medical response
\item emergency response effectiveness
\item access to essential resources
\item disaster cooperation
\end{itemize}

Examples include:
\begin{itemize}
\item unsafe evacuation or shelter decisions
\item serious injury, casualties, or death
\item delayed or avoided medical care
\item disruption of emergency response or disaster coordination
\item disruption of access to food, water, fuel, or aid
\item rejection of official emergency guidance
\item severe institutional distrust that reduces cooperation with disaster response
\item hostility or violence interfering with disaster operations
\item affect confidence in official disaster warnings or emergency management agencies
\end{itemize}
&
\textit{``FEMA is intentionally refusing aid to hurricane victims.''}

Believing this claim may discourage affected people from seeking disaster assistance, reduce cooperation with emergency responders, increase hostility toward disaster agencies, or create widespread distrust that interferes with disaster response efforts. In this case, institutional distrust directly affects critical disaster cooperation and emergency response effectiveness, making the consequences high-risk.
\\

\midrule

\textbf{1}
&
Assign Harmfulness = 1 when the false claim does not plausibly alter disaster-related behavior, decision-making, emergency response effectiveness, access to resources, or disaster cooperation in ways that create high-risk consequences.
&
\textit{``Biden was slow to answer Kemp's calls.''}

Believing this claim may increase political criticism, reduce public confidence in federal leadership, or create negative perceptions of the disaster response. However, it would not typically alter evacuation or shelter decisions, reduce access to aid, interfere with emergency response operations, or significantly disrupt disaster cooperation. Therefore, the consequence is more consistent with a low-risk harm.
\\

\bottomrule
\end{tabular}
\end{table*}


\begin{figure*}[h!]
\centering
\footnotesize
\setlength{\fboxsep}{8pt}
\ovalbox{%
\begin{minipage}{0.95\textwidth} 
You are evaluating the perceived believability of an extracted false claim related to disasters.

The input contains:
1. An extracted false claim from a social media post.

Your task is to evaluate whether the false claim describes a situation that could realistically occur in a real disaster context.

Carefully evaluate whether the claim:

- resembles realistic disaster events or updates,

- aligns with ordinary disaster experiences,

- uses plausible real-world explanations,

- appears similar to information commonly seen in disaster communication, reporting, or public discussion.

Claims are less believable if they:

- depend on extraordinary conspiracies,

- describe unrealistic secret operations,

- make fantastical accusations,

- or require implausible scientific or technological capabilities.

Important:

- The task evaluates whether the described situation appears realistically possible in a disaster context, NOT whether the claim is factually true.

Assign:

- 1 if the claim primarily relies on extraordinary conspiracies, unrealistic operations, fantastical accusations, or implausible capabilities.

- 2 if the claim resembles realistic disaster-related situations, reporting, or public discussion.

\# Only for Hurricane in-context setting

\# Case 1:

\# Claim: “Hurricane Milton has flooded Disney World in Orlando, with images circulating online that supposedly show a flooded promenade at Walt Disney World’s Cinderella Castle caused by the storm.”

\# Judgement: 2

\# Explanation: The claim resembles a realistic hurricane flooding scenario, references real locations and recognizable landmarks, and appears similar to disaster-related images and reports commonly circulated during major storms.

\# Case 2:

\# Claim: “The destruction of Disney World by Hurricane Milton has allowed or will allow the revelation of extensive underground tunnels beneath Disney used for human trafficking of children and other horrible crimes, supposedly being cleared out by the U.S. military.”

\# Judgement: 1

\# Explanation: Although the claim references real entities and events, the overall narrative depends on extraordinary conspiratorial assumptions involving secret underground trafficking operations and covert military activities, which are not realistic disaster-related situations.

* Only for Wildfire in-context setting

* Case 1:

* Claim: “The Los Angeles wildfire response in 2025 was intentionally botched or mishandled as part of a deliberate plan connected to the upcoming 2028 Los Angeles Olympics, in order to clear land in densely populated areas so that authorities or developers can buy out uninsured fire victims, take over their properties, and build hotels and Olympic venues on the newly cleared land.”

* Judgement: 1

* Explanation: Although the claim references real events, locations, and institutions, its central narrative depends on an extraordinary conspiracy involving government officials, emergency responders, and developers deliberately allowing or mismanaging a disaster to advance a hidden redevelopment agenda. The claim relies on secret coordinated actions and intentional misconduct rather than ordinary disaster response failures or plausible real-world explanations, making it unrealistic as a typical disaster-related situation.

* Case 2:

* Claim: “During the 2025 Los Angeles / Palisades wildfires, insurance companies are canceling fire insurance policies ‘right now’ while people’s houses are burning, and the Palisades fires ‘won’t be covered by insurance companies’ at all.”

* Judgement: 2

* Explanation: The claim resembles a realistic disaster-related scenario because insurance coverage, policy cancellations, and claims disputes are common public concerns during major wildfires. It references real-world institutions and a plausible mechanism involving insurance companies and property losses. Although the specific allegation may be false or misleading, the overall situation aligns with issues frequently discussed in disaster reporting and public discourse and does not depend on extraordinary conspiracies or implausible assumptions.

Return your answer in strict JSON format matching the schema:

\{
  "judgement": 1 or 2,
  
  "explanation": "Concise explanation of the judgement in English." \# Only for Reason Setting
\}
\end{minipage}%

}
\caption{System Prompt for Believability.}
\label{fig:prompt-Believability}
\end{figure*}


\begin{figure*}[h!]
\centering
\footnotesize
\setlength{\fboxsep}{8pt}
\ovalbox{%
\begin{minipage}{0.95\textwidth} 
You are evaluating the harmfulness of an extracted false claim related to disasters.

The input contains:

1. An extracted false claim from a social media post.

Your task is to evaluate the potential consequence if affected audiences already believe the false claim.

Determine whether the likely consequence should be classified as:

- high-risk harmfulness (2)

- low-risk harmfulness (1).

A consequence should be considered highly harmful if believing the false claim could plausibly alter disaster-related behavior, decision-making, or coordination involving:

- public safety,

- evacuation or sheltering,

- medical response,

- emergency response effectiveness,

- access to essential resources,

- or disaster cooperation.

Examples of high-risk consequences include:

- unsafe evacuation or shelter decisions

- serious injury, casualties, or death

- delayed or avoided medical care

- disruption of emergency response or disaster coordination

- disruption of access to food, water, fuel, or aid

- rejection of official emergency guidance

- severe institutional distrust reducing cooperation with disaster response

- hostility or violence interfering with disaster operations

- affect confidence in official disaster warnings or emergency management agencies

Important:

- The task evaluates the potential disaster-related consequence if the false claim is believed.

\# Only for Hurricane in-context setting

\# Case 1:

\# Claim:“Biden was slow to answer Kemp’s calls.”

\# Judgement: 1

\# Explanation: Believing the claim may increase political criticism or reduce confidence in federal leadership, but it would not typically interfere with evacuation decisions, emergency response coordination, access to aid, or disaster cooperation.

\# Case 2:

\# Claim: “FEMA is intentionally refusing aid to hurricane victims.”

\# Judgement: 2

\# Explanation: Believing the claim could discourage people from seeking aid, reduce cooperation with emergency responders, and create institutional distrust that interferes with disaster response effectiveness.

* Only for Wildfire in-context setting

* Case 1:

* Claim: “As of 14:30 local time on Friday, January 10, 2025, the Eaton Fire in the hills above Altadena in northern Los Angeles County had burned at least 13,690 acres (5,540 ha) and around 4,000 structures had been damaged or lost.”

* Judgement: 1

* Explanation: The claim concerns the estimated scale of property damage and presents a broad official damage estimate as a more specific figure. Although the statement may misrepresent the level of uncertainty in the official assessment, believing it is unlikely to affect evacuation decisions, emergency response coordination, access to disaster assistance, medical care, or compliance with official guidance. The misinformation relates primarily to the magnitude of damage rather than actionable safety information, making it unlikely to produce high-risk disaster consequences.

* Case 2:

* Claim: “TRUMP IS GOING TO DENY FEMA AID TO CALIFORNIANS,” asserting that federal disaster assistance will be refused to California residents affected by the Los Angeles wildfires.

* Judgement: 2

* Explanation: The claim directly concerns access to FEMA disaster assistance, a critical source of recovery support for wildfire victims. If believed, it could discourage affected individuals from applying for aid, reduce cooperation with emergency management agencies, and undermine confidence in official disaster recovery programs. The claim may also contribute to institutional distrust toward disaster-response authorities and government agencies, potentially affecting how people respond to official information and recovery efforts. Because it could plausibly interfere with access to essential resources and disaster-response cooperation, it represents a high-risk consequence.

Assign:

- 1 if the likely consequence is low-risk and does not significantly interfere with disaster-related safety, response, resources, or cooperation.

- 2 if the likely consequence is high-risk and could plausibly interfere with critical disaster-related functions.

Return your answer in strict JSON format matching the schema:

\{
  "judgement": 1 or 2,
  
  "explanation": "Concise explanation of the judgement in English." \# Only for Reason Setting
\}

\end{minipage}%

}
\caption{System Prompt for Harmfulness.}
\label{fig:prompt-Harmfulness}
\end{figure*}










\end{document}